%% file: spc_mps.tex
\documentclass[aps,prb,twocolumn,superscriptaddress,nofootinbib,longbibliography,floatfix]{revtex4-2}
\input{preamble}

\begin{document}

\title{Structure-preserving compression of Matrix Product States}

\author{Pouriya Ghadimi\,\orcidlink{0009-0000-0101-3653}}
\affiliation{Independent Researcher, Eindhoven, Netherlands}

\author{Maximilian Fröhlich\,\orcidlink{0009-0007-5276-2858}}
\affiliation{Weierstrass Institute for Applied Analysis and Stochastics, Berlin, Germany}

\author{Aaron Sander\,\orcidlink{0009-0007-9166-6113}}
\affiliation{Technical University of Munich, Munich, Germany}

\author{Martin Eigel\,\orcidlink{0000-0003-2687-4497}}
\affiliation{Weierstrass Institute for Applied Analysis and Stochastics, Berlin, Germany}

\author{Richard M.~Milbradt\,\orcidlink{0000-0001-8630-9356}}
\affiliation{Technical University of Munich, Munich, Germany}

\date{November 17, 2026}

\begin{abstract}
Matrix product state time evolution relies on compression to
control the growth of bond dimensions, but compression can break the
conservation laws of the equation being solved. A truncation is not
the flow of a Hermitian generator, so even a small truncation error
can violate exact algebraic identities of the Schr\"odinger equation.
For any matrix product state integrator that leaves the state in canonical form with
an orthogonality center, a scalar correction restores an admissible
global invariant after compression whenever the resulting scalar quadratic constraint equation admits a real root, without enlarging any bond dimension.
Rank-preserving recovery searches other canonical centers when
that correction fails. The geometry of the low-rank manifold provides sufficient conditions
for conservation by the fixed-rank projected flow and bounds what
corrections within a frozen center frame can achieve. Numerical experiments demonstrate that
recovery can be required even along accurate truncated trajectories.
\end{abstract}

\maketitle

\input{sections/introduction}
\input{sections/related_work_section}
\input{sections/preliminaries}
\input{sections/correction}
\input{sections/guarantees}
\input{sections/admissibility}
\input{sections/joint}
\input{sections/numerics}

\input{sections/conclusion}

\section*{Acknowledgments}
This research was supported by the Einstein Research Unit (ERU) on quantum devices at WIAS and conducted jointly with the Technical University of Munich. Martin Eigel acknowledges partial financial support from the German Federal Ministry of Education and Research (BMBF) under grant agreement No.~13N17160 (Q-ROM - Quantum Read-Once-Memory: Verwandlung von klassischen Daten zu Quantenzuständen). It has also been funded by the European Union under the Horizon Europe Programme by the European Research Council projects DA QC (Grant Agreement 101001318) and is part of the Munich Quantum Valley, which is supported by the Bavarian state government with funds from the Hightech Agenda Bayern Plus. The authors acknowledge the use of AI-based tools to support the preparation of this manuscript. All AI-assisted output was critically reviewed and verified by the authors. The authors assume responsibility for all content.

\bibliography{references}
\appendix
\input{appendices}

\bibliographystyle{apsrev4-2}

\end{document}

%% file: preamble.tex
\usepackage{amsmath,amssymb,amsfonts,mathtools}
\usepackage{amsthm}
\usepackage{bm}
\usepackage{graphicx}
\usepackage{tikz}
\usepackage{booktabs}
\usepackage{algpseudocode}
\usepackage{orcidlink}

\newcounter{algctr}
\renewcommand{\thealgctr}{\arabic{algctr}}
\newenvironment{algorithm}[1][t]{%
  \begin{figure}[#1]%
  \refstepcounter{algctr}%
  \hrule height 0.8pt\vspace{2pt}%
}{%
  \vspace{1pt}\hrule height 0.8pt%
  \end{figure}%
}
\newcommand{\algcaption}[1]{%
  \noindent\textbf{Algorithm \thealgctr.}\ #1\par
  \vspace{2pt}\hrule height 0.4pt\vspace{3pt}}

\newtheorem{theorem}{Theorem}
\newtheorem{proposition}[theorem]{Proposition}
\newtheorem{lemma}[theorem]{Lemma}

\theoremstyle{definition}

\theoremstyle{remark}

\algrenewcommand{\algorithmiccomment}[1]{\hfill$\triangleright$ #1}
\algrenewcommand\algorithmicrequire{\textbf{Input:}}
\algrenewcommand\algorithmicensure{\textbf{Output:}}

\newcommand{\cH}{\mathcal{H}}                 
\newcommand{\cL}{\mathcal{L}}                 
\newcommand{\cR}{\mathcal{R}}                 
\newcommand{\M}{\mathcal{M}}                  
\newcommand{\C}{\mathbb{C}}
\newcommand{\R}{\mathbb{R}}
\newcommand{\Heff}{H_{\mathrm{eff}}}
\newcommand{\Oeff}{O_{\mathrm{eff}}}
\newcommand{\Id}{\mathrm{Id}}                
\newcommand{\F}{\mathrm{F}}                   
\newcommand{\dd}{\mathrm{d}}
\newcommand{\Cov}{\mathrm{Cov}}

\newcommand{\ran}{\mathrm{ran}}


%% file: sections/introduction.tex

\section{Introduction}
\label{sec:intro}

The bond dimensions required to represent a quantum state as a Matrix product state (MPS) grow with the
entanglement generated by the dynamics, and a practical calculation must compress
the state by discarding singular values. The
compression makes the method affordable and can break the conservation laws of the
underlying equation. A truncated singular value
decomposition is not the flow of any Hermitian generator, so wherever the bond
dimension is reduced the invariants of the dynamics can be lost. Along the exact flow of
$\dot\psi=-i\hat H\psi$ the identity
$\langle\psi(t)|\hat H|\psi(t)\rangle=\langle\psi(0)|\hat H|\psi(0)\rangle$
follows algebraically from $\hat H=\hat H^\dagger$ and is not an error bound.

The one-site time-dependent
variational principle (TDVP) conserves the norm and, for a time-independent
Hamiltonian, the energy exactly~\cite{Haegeman2016}, because its update is
generated by a Hermitian effective operator at every site. The two-site variant,
which is used when the rank must adapt, truncates inside each local update and
loses this property. Rank-adaptive basis-update and Galerkin (BUG)
integrators~\cite{Ceruti2022,Ceruti2024} are in the same position. Their
uncompressed sweep leaves the norm and the energy unchanged to the accuracy of
its own time discretization, and the compression that follows it does not.

The classical remedy for a one-step method that leaves the level set of a first
integral is the projection method~\cite[\S IV.4]{HairerLubichWanner2006}, which
displaces the numerical solution along constraint gradients until the
constraint is met. In Hilbert space, however, the energy gradient at
$|\psi\rangle$ is proportional to $(\hat H-E\Id)|\psi\rangle$, with $E$ the
current expectation value. Applying the matrix product operator (MPO)
$\hat H$ can enlarge the MPS bond dimensions, so the displacement generally
leaves the fixed-bond-dimension manifold~\cite{Holtz2012,Uschmajew2013}.
Recompression can again disturb the invariant. The correction should therefore
act within the compressed representation. Here ``projection method'' refers
to this correction of first integrals, rather than to a low-rank
time integrator~\cite{Kieri2018}.

We construct such a correction at the orthogonality center. With the
surrounding canonical tensors fixed, the expectation value of a Hermitian MPO
is a Rayleigh quotient of an effective operator on the single center tensor.
Along its gradient direction, the exact target constraint reduces to a scalar
quadratic. A real root restores the expectation value with two matrix-free
effective-operator applications and no bond dimension changed
(Sec.~\ref{sec:correction}). If the scalar solve fails, rank-preserving recovery
tries other canonical centers and then the complete one-site frame.
Section~\ref{sec:guarantees} establishes the guarantees and limits of correction
within a fixed frame, and Sec.~\ref{sec:joint} treats several invariants together.

The choice of target also matters. Section~\ref{sec:admissibility} derives
sufficient conditions for conservation by the fixed-rank projected flow,
covering energy, conserved on-site sums, and their real linear combinations.
Commutation with the Hamiltonian alone does not guarantee this property;
quantities such as $\hat S^2$ require a separate assessment. Conservation by a
particular uncompressed numerical update must also be checked separately.
Section~\ref{sec:numerics} demonstrates joint restoration for BUG and two-site
TDVP, and scalar recovery along accurate truncated trajectories.

%% file: sections/related_work_section.tex
\section{Related work}
\label{sec:related-work}

\subsection{Projection and conservative low-rank methods}

Projection methods displace a provisional solution along constraint directions
and determine multipliers from prescribed invariant values
\cite{NortonEtAl2013Projection}. Under the usual nondegeneracy assumptions,
they retain the order of the underlying method. Relaxation Runge--Kutta
methods instead rescale the step to conserve norms
\cite{Ketcheson2019Relaxation}, Hamiltonian energies
\cite{RanochaKetcheson2020Hamiltonian}, or several nonlinear quantities
\cite{BiswasKetcheson2023MultipleRelaxation}. Correction along gradients and
scalar or coupled multiplier problems are therefore classical.

In tensor-train methods, linear invariants can be retained by expanding the
approximation subspace with their generating vectors
\cite{Dolgov2019Conservation}. Conservative low-rank schemes for kinetic
equations build corrections from the factors, modify the Galerkin condition,
or protect moment-carrying subspaces during truncation
\cite{EinkemmerLubich2019,EinkemmerJoseph2021Conservative,GuoQiu2022ConservativeTensor}.
Rank-adaptive variants combine basis augmentation with conservative truncation
\cite{EinkemmerOstermannScalone2022Robust}, and conservative BUG integrators
retain conserved basis functions in the evolved basis~\cite{Einkemmer2023}.
These schemes target linear moments and can retain local balance laws. Here
the targets are global quantum expectation values, hence ratios of quadratic
forms; the preceding compression is unchanged and the correction is applied
afterwards.

\subsection{MPS and tensor-network methods}

Mixed-canonical form, effective operators, and the MPS tangent-space projector
are standard~\cite{Schollwock2011,Haegeman2011}. For a time-independent
Hamiltonian, one-site TDVP conserves norm and energy at fixed rank;
truncation in bond-adaptive variants loses this
property~\cite{Haegeman2016}. Rank-adaptive BUG integrators update the bases,
perform a Galerkin step, and truncate~\cite{Ceruti2022}; their canonical-MPS
realization uses sweeps~\cite{Frohlich2026}.
Kap\'{a}s et al.~\cite{KapasEtAl2026BUG} formulate augmented and midpoint BUG
integrators for MPS with an interior-bond Galerkin update and benchmark them
against projector-splitting TDVP. Their schemes preserve norm and energy in
the absence of truncation. The correction below is appended to compression
and leaves these updates unchanged.

Other approaches alter the representation or truncation: Hamiltonian
augmentation conserves the average energy of an evolved MPO~\cite{Mendl2018};
variational truncation improves a lower-rank MPS approximation
\cite{VanheckeEtAl2020Truncating}; controlled bond expansion reduces TDVP
projection error~\cite{LiGleisVonDelft2024CBE}; and goal-oriented CP and Tucker
decompositions incorporate selected quantities in the low-rank fit
\cite{DunlavyEtAl2025GoalOriented}. None gives an exact post-compression
expectation correction at unchanged MPS bond dimensions.

Symmetry-adapted tensors preserve Abelian charge sectors
\cite{SinghPfeiferVidal2011U1} or non-Abelian multiplets
\cite{SinghVidal2012SU2,Weichselbaum2012NonAbelian}, which is stronger than
fixing expectation values. Our correction requires no symmetry-labelled
tensors and restores only its specified expectations. The statements about
$\langle\hat S^2\rangle$ concern the generic MPS manifold.

Here the classical projection construction acts inside the frozen canonical
frame of a compressed MPS, where the exact scalar constraint is quadratic.
The following analysis establishes its rank-preserving guarantees and limits.

%% file: sections/preliminaries.tex

\section{Canonical form, effective observables, and compression}
\label{sec:prelim}

\subsection{Matrix product states and the orthogonality center}
\label{sec:prelim_mps}

Let $\cH=\bigotimes_{k=1}^{L}\C^{d_k}$ be the state space of an $L$-site
chain. An MPS of bond dimensions $\bm r=(r_1,\dots,r_{L-1})$ represents
$|\psi\rangle\in\cH$ (also written $\psi$) by the contraction of site tensors
$A^{[k]}\in\C^{r_{k-1}\times r_k\times d_k}$ with $r_0=r_L=1$. The
representation is not unique, since inserting $X X^{-1}$ on a bond and absorbing
the factors into the neighbouring tensors leaves $|\psi\rangle$ unchanged. This
gauge freedom is fixed by the mixed-canonical form
\begin{equation}\label{eq:mixed_canonical}
  |\psi\rangle
  \;=\;
  L^{[1]}\cdots L^{[k-1]}\; C^{[k]}\; R^{[k+1]}\cdots R^{[L]},
\end{equation}
in which every tensor to the left of site $k$ is left-isometric, every tensor to
its right is right-isometric, and the single remaining non-isometric tensor
$C^{[k]}$ is the \emph{orthogonality center} (OC). Any MPS can be brought to
this form by a sequence of QR decompositions, and the center can be moved to a
neighbouring site by one further QR.

Holding the isometric factors of Eq.~\eqref{eq:mixed_canonical} fixed, each
tensor $C$ of the shape of $C^{[k]}$ defines a state through the linear
\emph{inclusion map} $J:\C^{r_{k-1}\times r_k\times d_k}\to\cH$,
\begin{equation}\label{eq:inclusion}
  J C \;:=\; L^{[1]}\cdots L^{[k-1]}\, C\, R^{[k+1]}\cdots R^{[L]},
\end{equation}
so that $|\psi\rangle=JC^{[k]}$.
The map $J$ is fixed during each center correction; moving the center
generally changes $J$ and $\ran J$ while leaving the represented state unchanged.
Because the surrounding tensors are isometric,
contracting $(JC)^\dagger(JC')$ collapses them pairwise onto identities and
leaves the Frobenius inner product of the two center tensors. Hence $J$ is an
isometry,
\begin{equation}\label{eq:isometry}
  J^\dagger J \;=\; \Id,
  \qquad
  \langle JC,\,JC'\rangle \;=\; \langle C,\,C'\rangle_\F .
\end{equation}
Here and below, $\Id$ denotes the identity on the indicated space.
The entire norm of the many-body state is therefore carried by the
single center tensor, $\||\psi\rangle\|=\|C^{[k]}\|_\F$. We write $JX$ for the
state obtained by exchanging the center for an arbitrary tensor $X$ of the same
shape while every other tensor is kept fixed.

\subsection{Effective observables and the Rayleigh identity}
\label{sec:prelim_eff}

For a nonzero state, we write
$\langle\hat O\rangle_\psi=\langle\psi|\hat O|\psi\rangle/\langle\psi|\psi\rangle$
for its normalized expectation, omitting the state subscript when clear.
Let $\hat O$ be a Hermitian MPO
\begin{equation}
    \hat{O} = W^{[1]} \cdots W^{[L]}.
\end{equation}
The quantity a correction must control is
$\langle\hat O\rangle_\psi$, which can involve all $L$ sites, while the
only tensor it may move is $C$. Compressing $\hat O$ by the isometry $J$,
\begin{equation}\label{eq:oeff}
  \Oeff \;:=\; J^\dagger \hat O\, J
  \;=\; \cL_k\, W^{[k]}\, \cR_k ,
\end{equation}
removes the rest of the chain from that constraint. We call $\Oeff$ the
\emph{effective observable} at the center. The second equality is the standard
environment form, in which $\cL_k$ and $\cR_k$ are the partial contractions of
the MPO with the isometric tensors to the left and to the right of site $k$, and
$W^{[k]}$ is the local MPO tensor. Neither depends on $C$, so both are
assembled once for a given center, at a cost of $\mathcal{O}(L)$ contractions,
and every subsequent evaluation of $\Oeff C$ is a fixed sequence of contractions
whose cost is independent of $L$~\cite[\S6.2]{Schollwock2011}. The operator $\Oeff$ inherits Hermiticity from
$\hat O$, which keeps the coefficients of the constraint of
Sec.~\ref{sec:correction_mu} real, and it is never formed explicitly. For $\hat
O=\hat H$ we write $\Heff$, the effective Hamiltonian that generates the local
update of every one-site integrator.

Equation~\eqref{eq:isometry} reduces the denominator to $\langle C,C\rangle_\F$
and Eq.~\eqref{eq:oeff} reduces the numerator to $\langle C,\Oeff C\rangle_\F$,
so the expectation value of $\hat O$ in the many-body state $JC$ is a Rayleigh
quotient of $\Oeff$ on the single tensor $C$,
\begin{equation}\label{eq:rayleigh}
  \frac{\langle JC|\hat O|JC\rangle}{\langle JC|JC\rangle}
  \;=\;
  R_O(C)
  \;:=\;
  \frac{\langle C,\,\Oeff\,C\rangle_\F}{\langle C,\,C\rangle_\F}.
\end{equation}
A correction is therefore affordable. A constraint on a vector with
$\prod_kd_k$ entries has become an exactly equivalent constraint on the
$r_{k-1}r_kd_k$ entries of a single tensor.

\subsection{Compression}
\label{sec:prelim_compression}

We write $\Pi_\vartheta$ for a compression sweep at relative tolerance
$\vartheta$ under a fixed bond cap. It traverses the chain once, truncating each
bond in turn with the orthogonality center adjacent to it, and leaves the center
at the endpoint where the traversal ends. Retained singular values are not
rescaled. The traversal is that of the compression pass of tensor-train
rounding~\cite[Alg.~2]{Oseledets2011}.

Two properties of the sweep are used in Sec.~\ref{sec:guarantees}. With the
retained frames fixed, write $\Pi_1,\dots,\Pi_{L-1}$ for the orthogonal
projectors on $\cH$ representing the bond truncations in the order the traversal
performs them. The singular value decomposition at each bond is taken in the
canonical gauge, so the retained and discarded Schmidt vectors are orthonormal
in $\cH$, and each $\Pi_m$ acts on the state as an orthogonal projector, the
frame projection of Ref.~\cite[Def.~3.9]{Grasedyck2010}. Each truncation selects
its retained basis inside the basis its predecessor retained, tensored with the
local space, so the retained subspaces are nested,
\begin{equation}\label{eq:nesting}
  \ran\Pi_{L-1}\subseteq\cdots\subseteq\ran\Pi_1 .
\end{equation}

Neither property depends on which integrator produced the state that
$\Pi_\vartheta$ acts on, nor on how many compressions a step performs.

%% file: sections/correction.tex

\section{The correction at the orthogonality center}
\label{sec:correction}

Let $\hat O$ be a Hermitian MPO that commutes with $\hat H$, and let $o_0$ be
its expectation value in the initial state, recorded once at $t_0$ through
Eq.~\eqref{eq:rayleigh}. For energy conservation, $\hat O=\hat H$. After a
compression the state is $|\psi\rangle=JC$ with $C$ the center tensor at the
endpoint the sweep left, and $R_O(C)$ has in general moved away from $o_0$. The
construction below needs only that $\hat O$ is Hermitian. Whether a given
$\langle\hat O\rangle$ is a quantity worth restoring is a separate question,
and commuting with $\hat H$ turns out not to settle it
(Sec.~\ref{sec:admissibility}).

\subsection{The multiplier equation}
\label{sec:correction_mu}

Write $E=R_O(C)$ for the post-compression value and set
\begin{equation}\label{eq:gradient}
  g \;=\; (\Oeff - E\,\Id)\,C,
  \qquad
  \langle C,\,g\rangle_\F \;=\; 0 ,
\end{equation}
the orthogonality being immediate from the definition of $E$. Up to a positive
scalar, $g$ is the gradient of the Rayleigh quotient. The first-order change of
$R_O$ under a displacement $\delta C$ is
$2\,\mathrm{Re}\langle\delta C,g\rangle_\F/\|C\|_\F^2$, so the change in the
invariant is the projection of the displacement onto $g$. For a normalized
center tensor $\|g\|_\F$ is the standard deviation of $\Oeff$ in $C$, and it
vanishes exactly when $C$ is an eigenvector of $\Oeff$.

For $g\neq0$, we choose $g$ as the correction direction because the minimum-norm
displacement satisfying the linearized constraint is a real multiple
of $g$. This first-order optimality motivates the direction; it does
not imply that the finite correction minimizes the displacement
subject to the exact constraint. 

We restrict the correction to the one-parameter family
\begin{equation}\label{eq:update}
  C \;\longmapsto\; C + \mu\,g,
  \qquad \mu\in\R,
\end{equation}
which is what the projection method for first
integrals~\cite[\S IV.4]{HairerLubichWanner2006} prescribes, with $g$ in the
role of the constraint gradient. We determine $\mu$ from the exact constraint
$R_O(C+\mu g)=o_0$ and not from the linearization that fixed the direction. With the surrounding tensors fixed, $\Oeff$ is a fixed linear operator.
Both $\langle C+\mu g,\Oeff(C+\mu g)\rangle_\F$ and
$\|C+\mu g\|_\F^2$ are polynomials of degree at most two in $\mu$,
so clearing the denominator gives an equation of degree at most two. Hermiticity of $\Oeff$ and $\langle C,g\rangle_\F=0$ then
fix its coefficients, and the constraint reduces exactly to the scalar quadratic
\input{figure_geometry}

\begin{equation}\label{eq:quadratic}
  a\mu^2+b\mu+c=0,\quad
  \begin{aligned}
    a &= \langle g,\Oeff\,g\rangle_\F - o_0\|g\|_\F^2,\\
    b &= 2\|g\|_\F^2,\\
    c &= \langle C,C\rangle_\F\,(E-o_0).
  \end{aligned}
\end{equation}
Equation~\eqref{eq:gradient} gives
$\langle g,\Oeff C\rangle_\F=\|g\|_\F^2$, so the linear coefficient costs no
additional contraction and $b>0$ whenever $g\neq0$.

Figure~\ref{fig:correction-geometry} illustrates compression and the scalar
correction within the fixed center frame.

The contrast with the general projection method is where the saving lies. There
the multiplier is determined by simplified Newton
iteration~\cite{HairerLubichWanner2006}, at a cost set by how fast that
iteration converges. Here the linearity of $\Oeff$ has already fixed the degree
at two, so one closed-form root replaces the iteration and the cost of the
correction is known in advance.

Two applications of $\Oeff$ suffice for the whole correction, one to evaluate
$E$ and one to form $\langle g,\Oeff g\rangle_\F$. Both use the environments of
Eq.~\eqref{eq:oeff}, so, if the integrator does not already have the required environments, the
correction requires $\mathcal{O}(L)$ contractions to assemble them and two
applications of $\Oeff$ thereafter. That is the order of the sweep the correction is
appended to, and not more. We denote
the real root of smallest modulus by $\mu_\star$. Since the displacement has
norm $|\mu|\|g\|_\F$, this root gives the smallest displacement along the
chosen line. It is
evaluated in the factored form
\begin{equation}\label{eq:citardauq}
  \mu_\star=\frac{c}{q},
  \qquad
  q=-\tfrac12\bigl(b+\operatorname{sign}(b)\sqrt{b^2-4ac}\,\bigr),
\end{equation}
which avoids cancellation as $a\to0$, the regime the correction operates in.
If $b^2<4ac$ the
level set does not intersect the line $\{C+\mu g\}$ and the state is returned
unchanged. At an eigenstate, $\hat O(JC)=\lambda(JC)$ implies $\Oeff C=\lambda C$
and $g=0$. If the target is already met within tolerance,
Algorithm~\ref{alg:correction} skips the scalar update before forming $g$.
If $g=0$ while the discrepancy exceeds tolerance, then $a=b=0$ and $c\neq0$,
so the scalar equation has no solution. This does not establish infeasibility
in the full frame; the recovery of Sec.~\ref{sec:correction_recovery} may still
restore the target.

The optional last line restores the norm. It is a separate and always admissible
correction. By Eq.~\eqref{eq:isometry} the center tensor carries the entire
norm of the many-body state, so the rescale $C\mapsto(\nu/\|C\|_\F)C$ sets
$\||\psi\rangle\|=\nu$ exactly. Because $R_O$ is scale invariant, the rescale
cannot disturb a restored invariant, and it is therefore applied last.

\subsection{Algorithm}
\label{sec:correction_alg}

\begin{algorithm}[tbp]
\algcaption{\textsc{RayleighCorrection}}
\label{alg:correction}
\begin{algorithmic}[1]
\Require canonical MPS $\psi=JC$ with center tensor $C$ at the endpoint left
         by compression; Hermitian MPO $\hat O$; target $o_0$; tolerance $\tau$;
         optional norm target $\nu$
\Ensure center tensor $C$, with optional norm rescaling
\Statex \textbf{Guarantees:} if the scalar update is accepted, $R_O(C)=o_0$
        up to round-off; otherwise it is skipped; all bond dimensions unchanged
\State $E \gets \langle C,\Oeff C\rangle_\F / \langle C,C\rangle_\F$
       \Comment{one application of $\Oeff$}
\If{$|E-o_0| \ge \tau\max(1,|o_0|)$}
       \Comment{else nothing to restore}
  \State $g \gets (\Oeff - E\,\Id)\,C$
  \State $a \gets \langle g,\Oeff g\rangle_\F - o_0\|g\|_\F^2$
       \Comment{second application of $\Oeff$}
  \State $b \gets 2\|g\|_\F^2$
  \State $c \gets \langle C,C\rangle_\F\,(E-o_0)$
  \If{$b > 0$ \textbf{and} $b^2 - 4ac \ge 0$}
       \Comment{else $g=0$ or no real root; leave $C$ undisplaced}
    \State $q \gets -\tfrac12\bigl(b+\operatorname{sign}(b)\sqrt{b^2-4ac}\,\bigr)$
    \State $\mu_\star \gets c/q$, $C \gets C + \mu_\star g$
  \EndIf
\EndIf
\If{$\nu$ supplied}
  \State $C \gets \bigl(\nu/\|C\|_\F\bigr)\,C$
       \Comment{$R_O$ is scale invariant}
\EndIf
\State \Return $C$
\end{algorithmic}
\end{algorithm}

In Algorithm~\ref{alg:correction}, the correction is applied to the center
that a completed compression leaves, before any renormalization. A scheme
with a separate compression sweep, such as the alternating-endpoint
composition of Ref.~\cite{Frohlich2026}, admits the correction inside each
compression map. Two-site TDVP truncates at every local two-site update;
for its symmetric sequence of local updates the correction is applied once per timestep at
the first site, where the center returns.

\subsection{Rank-preserving failure recovery}
\label{sec:correction_recovery}

Let $k_0$ denote the center left by compression. A rejected scalar solve of
Eq.~\eqref{eq:quadratic} for $\mu$ concerns only one direction at one center. The
orthogonality center is first shifted through the chain by exact QR gauge
transformations, repeating the scalar solve at each other site until one is
accepted. These shifts preserve the physical state and every bond dimension.
The value $E$ is unchanged until acceptance, while the effective operator and
correction direction depend on the center. The reference implementation
rebuilds environments at each center, requiring up to $\mathcal{O}(L^2)$ environment
contractions for a full scan at fixed bond dimensions.

If all scalar solves fail, the recovery returns to $k_0$ and tests the
complete one-site frame. At each center $k$ being tested, write $C^{[k]}$
and $\Oeff^{[k]}$ for its center tensor and effective observable, and define
$A_k=\Oeff^{[k]}$ and $B_k=A_k-o_0\Id$.
A nonzero tensor $X$ satisfying $R_O(X)=o_0$ exists in this frame if and only if
\begin{equation}
\label{eq:frame-feasibility}
  o_0\in\bigl[\lambda_{\min}(A_k),\lambda_{\max}(A_k)\bigr].
\end{equation}
Here $\lambda_{\min}(A_k)$ and $\lambda_{\max}(A_k)$ are the smallest and
largest eigenvalues of $A_k$. Since $E=R_O(C^{[k]})$ lies in this interval,
only the extremum on the
side of the target is needed, i.e.
\begin{equation}
\label{eq:frame-target-side}
\begin{aligned}
  \lambda_{\min}(A_k)\leq o_0 & \text{ if } E>o_0,\\
  \lambda_{\max}(A_k)\geq o_0 & \text{ if } E<o_0.
\end{aligned}
\end{equation}
If condition (\ref{eq:frame-feasibility}) is not fulfilled at $k_0$, the sites are visited in order until
the first feasible one is found.

The extremal eigenvalue can be evaluated matrix-free by Lanczos iteration.
The starting directions must explore the admissible frame: a start using
$C^{[k]}$ alone can miss eigenspaces orthogonal to it. With spectral tolerance
$\tau_{\mathrm{sp}}$, a smallest Ritz value at most $o_0+\tau_{\mathrm{sp}}$
for $E>o_0$, or a largest Ritz value at least $o_0-\tau_{\mathrm{sp}}$ for
$E<o_0$, establishes feasibility to that tolerance. Failure of a Ritz value
to cross the target does not establish infeasibility; that conclusion
requires convergence of the extremal solve or a certified spectral bound.

At the first feasible center $k$ , the least-displacing update solves
\begin{equation}
\label{eq:frame-minimum}
\begin{aligned}
  X_k^\star &\in
  \operatorname*{arg\,min}_{X\neq0}\frac12\|X-C^{[k]}\|_\F^2,\\
  &\text{subject to }\langle X,B_kX\rangle_\F=0,
\end{aligned}
\end{equation}
provided the minimum is attained at a nonzero tensor. For a regular
minimizer, the stationarity equations are
\begin{equation}
\label{eq:frame-secular}
\begin{aligned}
  X_k(\lambda)&=(\Id+\lambda B_k)^{-1}C^{[k]},\\
  \phi_k(\lambda)&=\langle X_k(\lambda),B_kX_k(\lambda)\rangle_\F=0.
\end{aligned}
\end{equation}
The minimum is selected on the branch $\Id+\lambda B_k\succeq0$; a regular
root $\lambda_\star$ on this branch gives $X_k^\star=X_k(\lambda_\star)$.
Indeed, for feasible $X_k^\star,Y$ and stationarity
$C^{[k]}=(\Id+\lambda B_k)X_k^\star$, setting $D=Y-X_k^\star$ gives
\begin{equation*}
\begin{gathered}
  \|Y-C^{[k]}\|_\F^2-\|X_k^\star-C^{[k]}\|_\F^2\\
  =\langle D,(\Id+\lambda B_k)D\rangle_\F\geq0,
\end{gathered}
\end{equation*}
which proves global minimality within this frame.
At a singular endpoint, the inverse is replaced by a pseudoinverse and a
component in its kernel is chosen to meet the constraint. At a boundary
of the spectral interval, the feasible tensors lie in $\ker B_k$ and the
minimizer is the orthogonal projection of $C^{[k]}$ onto that space. If this
projection vanishes, the nonzero-constrained problem has an unattained
infimum, so no minimizing update is accepted. Regular roots are bracketed and solved by bisection, using shifted
matrix-free conjugate-gradient solves at each trial. The positive-definite
search interval is estimated from the extremal eigenvalue computed in the
feasibility test, positive semidefiniteness is not rigorously certified.

Before any optional norm rescaling, the update minimizes displacement within
the selected one-site frame. The distance guarantee is restricted to the
compression frame (Sec.~\ref{sec:guarantees}). An update is accepted only after
its Rayleigh quotient meets the original tolerance; we then set
$C^{[k]}\gets X_k^\star$ and apply any optional norm rescaling.
If no center is feasible, exact restoration by a single-tensor update is
impossible and the compressed state is retained. A failed numerical solve
or an unattained minimum also leaves that state undisplaced, but is distinct
from spectral infeasibility. All center shifts and both recovery stages
preserve every bond dimension.

%% file: figure_geometry.tex
\begin{figure*}[t]
\centering
\begingroup
\definecolor{spcink}{HTML}{000000}
\definecolor{spcmuted}{HTML}{666666}
\definecolor{spcblue}{HTML}{0B5394}
\definecolor{spcplane}{HTML}{D9D9D9}
\definecolor{spcred}{HTML}{C1272D}
\resizebox{\textwidth}{!}{%
\begin{tikzpicture}[x=2.6cm,y=2.6cm,>=stealth,font=\small,
  line cap=round,line join=round,text=spcink,
  note/.style={font=\footnotesize},
  leader/.style={line width=.35pt},
  correction/.style={->,draw=spcblue,line width=1.1pt}]
\path[use as bounding box] (-.95,-1.08) rectangle (5.55,1.48);
\node[anchor=west] at (-.90,1.36) {(a) Compression, then correction};
\node[anchor=west] at (2.53,1.36) {(b) The quadratic selects the correction};

\filldraw[fill=spcplane,fill opacity=.40,draw=spcmuted,line width=.45pt]
  (.5375,-.4375) -- (1.8375,-.4375) -- (.4625,.4375)
  -- (-.8375,.4375) -- cycle;
\foreach \r in {.38,.75,1.12} {
  \draw[spcred,opacity=.25,line width=.4pt,dash pattern=on 2pt off 1.5pt]
    plot[domain=180:360,samples=65,variable=\t]
    ({\r-.55*\r*cos(\t)},{.35*\r*cos(\t)+\r*sin(\t)/sqrt(2.272189349)});
  \draw[spcred,opacity=.40,line width=.4pt]
    plot[domain=0:180,samples=65,variable=\t]
    ({\r-.55*\r*cos(\t)},{.35*\r*cos(\t)+\r*sin(\t)/sqrt(2.272189349)});
}
\draw[spcred,line width=.8pt]
  plot[domain=0:180,samples=65,variable=\t]
  ({1.12-.616*cos(\t)},{.392*cos(\t)+1.12*sin(\t)/sqrt(2.272189349)});
\draw[spcred,opacity=.65,line width=.6pt,dash pattern=on 2pt off 1.5pt]
  plot[domain=180:360,samples=65,variable=\t]
  ({1.12-.616*cos(\t)},{.392*cos(\t)+1.12*sin(\t)/sqrt(2.272189349)});
\foreach \t in {45,90,135} {
  \draw[spcred,opacity=.22,line width=.35pt] (0,0) --
    ({1.12-.616*cos(\t)},{.392*cos(\t)+1.12*sin(\t)/sqrt(2.272189349)});
}
\foreach \t in {225,270,315} {
  \draw[spcred,opacity=.18,line width=.35pt,dash pattern=on 2pt off 1.5pt]
    (0,0) -- ({1.12-.616*cos(\t)},{.392*cos(\t)+1.12*sin(\t)/sqrt(2.272189349)});
}
\draw[spcred,line width=1pt] (.504,.392) -- (0,0) -- (1.736,-.392);
\draw[spcmuted,dashed,line width=.5pt] (0.735939,0.595367) -- (0.322490,0.250826);
\draw[->,spcink,line width=1pt] (0.735939,0.595367) -- (0.735939,0.057883);
\draw[correction] (0.735939,0.057883) -- (0.322490,0.250826);
\draw[spcink,line width=.4pt] (0.735939,0.132883) -- (0.680782,0.158623) -- (0.680782,0.083623);
\fill[spcink] (0.735939,0.595367) circle[radius=1.6pt];
\fill[spcink] (0.735939,0.057883) circle[radius=1.6pt];
\fill[spcblue] (0.322490,0.293826) -- (0.311622,0.265785) -- (0.281595,0.264114) -- (0.304905,0.245112) -- (0.297216,0.216038) -- (0.322490,0.232336) -- (0.347765,0.216038) -- (0.340075,0.245112) -- (0.363386,0.264114) -- (0.333358,0.265785) -- cycle;
\draw[leader,spcmuted] (0.735939,0.595367) -- (.81,.88);
\draw[leader,spcmuted] (0.735939,0.057883) -- (.86,-.015);
\draw[leader,spcblue,->,shorten >=2pt] (.08,.26) -- (0.322490,0.250826);
\node[anchor=west,fill=white,inner sep=1pt] at (.82,.91) {$|\hat\psi\rangle$};
\node[anchor=west,text=spcink,fill=white,inner sep=1pt]
  at (.88,-.035) {$|\psi\rangle=JC$};
\node[anchor=south east,text=spcblue] at (.06,.19) {$|\psi_{\mathrm{corr}}\rangle$};
\node[rotate=90,text=spcink,note] at (.84,.33) {compression};
\node[rotate=-25,text=spcblue,fill=white,inner sep=1pt]
  at (.46,.07) {$\mu_\star\,Jg$};
\draw[leader,spcred] (1.05,.66) -- (1.16,.94);
\node[anchor=west,text=spcred] at (1.10,1.02) {$\langle\hat O\rangle=o_0$};
\node[anchor=west,note] at (1.10,.88) {target cone surface};
\node[anchor=west,text=spcink] at (-.64,-.69) {$\mathcal V=\ran J$};
\node[anchor=west,note] at (-.64,-.82) {fixed canonical frame};
\draw[spcred,fill=white,line width=.55pt] (0,0) circle[radius=1.35pt];
\node[anchor=north west,note,text=spcmuted,inner sep=1pt] at (0,-.025) {$0$};

\begin{scope}[shift={(4.15,-.38)}]
\begin{scope}
\clip (-1.62,-.17) rectangle (1.25,1.46);
\draw[spcred,line width=1pt] (-0.078706,-0.118058) -- (1.011929,1.517893);
\draw[spcred,line width=1pt] (0.177088,-0.118058) -- (-2.276840,1.517893);
\end{scope}
\draw[->,spcmuted,line width=.4pt] (-1.57,0) -- (1.16,0);
\draw[->,spcmuted,line width=.4pt] (0,-.14) -- (0,1.33);
\node[anchor=north,text=spcink] at (1.08,-.025) {$g$};
\node[anchor=west] at (.045,1.30) {$C$};
\node[anchor=north east,note,text=spcmuted,inner sep=1pt] at (-.025,-.02) {$0$};
\draw[spcred,fill=white,line width=.55pt] (0,0) circle[radius=1.35pt];
\draw[spcblue,line width=.65pt,dash pattern=on 3pt off 2pt] (-1.56,0.843274) -- (1.16,0.843274);
\draw[correction] (-0.000000,0.843274) -- (0.562183,0.843274);
\draw[spcmuted,line width=.4pt] (.09,0.843274) -- (.09,0.753274) -- (0,0.753274);
\fill[spcink] (-0.000000,0.843274) circle[radius=1.6pt];
\fill[spcblue] (0.562183,0.886274) -- (0.551315,0.858233) -- (0.521287,0.856562) -- (0.544598,0.837560) -- (0.536908,0.808486) -- (0.562183,0.824784) -- (0.587457,0.808486) -- (0.579768,0.837560) -- (0.603078,0.856562) -- (0.573051,0.858233) -- cycle;
\draw[spcmuted,fill=white,line width=.7pt] (-1.264911,0.843274) circle[radius=1.8pt];
\draw[leader,spcmuted] (-1.264911,0.843274) -- (-1.34,1.12);
\draw[leader,spcred] (0.730838,1.096256) -- (.60,1.49);
\node[anchor=south west,note,text=spcmuted] at (-1.51,1.14) {other root};
\node[anchor=west,text=spcred] at (.23,1.56) {$R_O(X)=o_0$};
\node[anchor=north east,text=spcink] at (-.025,0.798274) {$C$};
\node[anchor=south,text=spcblue] at (.28,0.898274) {$\mu_\star g$};
\node[anchor=north west,text=spcblue] at (0.607183,0.783274) {$C+\mu_\star g$};
\node[anchor=west,text=spcblue] at (-.96,0.983274) {$\ell=\{C+\mu g\}$};
\node[anchor=west] at (-1.50,-.32) {$g\perp C$};
\node[anchor=west,note] at (-.49,-.32) {Choose the nearer intersection.};
\end{scope}
\end{tikzpicture}%
}
\endgroup
\caption{\textbf{Geometry of compression and the scalar correction} (illustrative real
slice, before norm rescaling). (a) Compression orthogonally projects
$|\hat\psi\rangle$ onto the fixed canonical frame
$\mathcal V=\ran J$. The red cone surface consists of nonzero $v$ satisfying
$\langle v|(\hat O-o_0\Id)|v\rangle=0$; the origin is excluded. One sheet is
shown, with dashed portions below the frame. Here $|\hat\psi\rangle$ lies on
the target surface, while $JC$ does not. Changing only $C$ moves within
$\mathcal V$, and the correction reaches the target surface. (b) The
intersections of $C+\mu g$ with the target set give the real roots of
Eq.~\eqref{eq:quadratic}. Here $\mu_\star$ denotes the real root of smallest
modulus, which minimizes the displacement along this line. The center
displacement is $\mu_\star g$; the corresponding state displacement is
$\mu_\star Jg$, so $|\psi_{\mathrm{corr}}\rangle=J(C+\mu_\star g)$.}
\label{fig:correction-geometry}
\end{figure*}
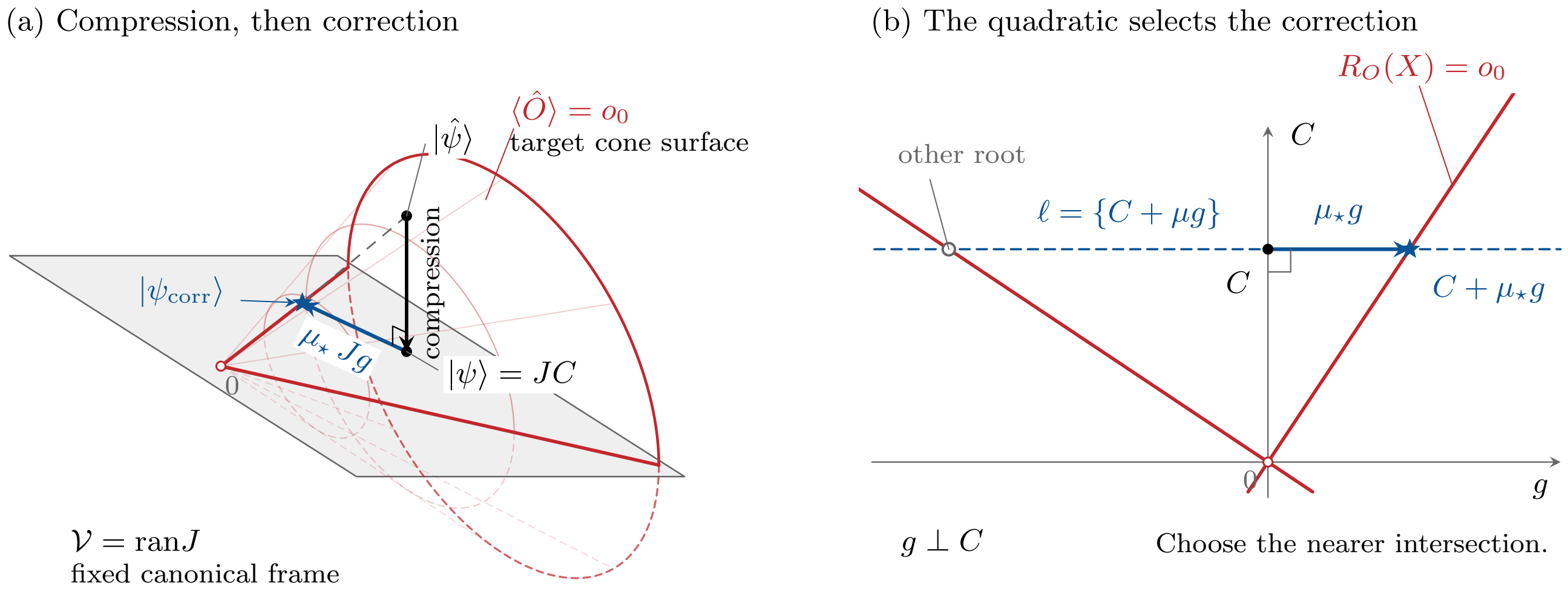

%% file: sections/guarantees.tex

\section{Guarantees and limits}
\label{sec:guarantees}

\begin{lemma}[The compression sweep is an orthogonal projection]
\label{lem:projection}
Let $|\hat\psi\rangle$ be the state before the compression sweep of
Sec.~\ref{sec:prelim_compression}, let $J$ be the inclusion
map~\eqref{eq:inclusion} built from the isometries the sweep leaves and from the
site at which it leaves the center, and let $C$ be the center it leaves. Then
\begin{equation}\label{eq:projection}
  C \;=\; J^\dagger|\hat\psi\rangle ,
  \qquad
  |\psi\rangle \;:=\; JC \;=\; Q\,|\hat\psi\rangle ,
  \qquad
  Q \;:=\; JJ^\dagger ,
\end{equation}
and $|\hat\psi\rangle-|\psi\rangle$ is orthogonal to every state of the form
$JX$.
\end{lemma}

Equation~\eqref{eq:projection} is the least-squares characterization of the
center, $C=\arg\min_X\|JX-\hat\psi\|$, in the form it takes for orthonormal
frames~\cite[Def.~2.4]{Grasedyck2010}. The projector $Q$ depends on the
input through the retained isometries.

\begin{proof}
Write the sweep as $|\psi\rangle=\Pi_{L-1}\cdots\Pi_1|\hat\psi\rangle$ and let
$|w_m\rangle\in\ran(\Id-\Pi_m)$ be the increment discarded by $\Pi_m$, so that
$|\hat\psi\rangle-|\psi\rangle=\sum_m|w_m\rangle$. The isometries the sweep
leaves span the subspace retained by its last truncation, so
$\ran J=\ran\Pi_{L-1}$, and the nesting~\eqref{eq:nesting} gives
$\ran J\subseteq\bigcap_m\ran\Pi_m$. Hence $\ran(\Id-\Pi_m)$ is orthogonal to
$\ran J$ and $\langle JX|w_m\rangle=0$ for every $m$ and every $X$. Applying
$J^\dagger$ to $|\hat\psi\rangle=JC+\sum_m|w_m\rangle$ and using
$J^\dagger J=\Id$ from Eq.~\eqref{eq:isometry} gives
$J^\dagger|\hat\psi\rangle=C$.
\end{proof}

Equation~\eqref{eq:nesting} collapses the composition
$\Pi_{L-1}\cdots\Pi_1$ to the single orthogonal projector $Q$. A composition of
orthogonal projections is in general not one, and without the nesting it need
not preserve the format~\cite[Ex.~3.14]{Grasedyck2010}. For an integrator that
truncates during local updates, the lemma applies when correction is performed
at the center left by the final truncation of a timestep. Taking
$|\hat\psi\rangle$ to be the state before that truncation gives a single
orthogonal projection, for which Eq.~\eqref{eq:nesting} holds trivially.

\begin{proposition}[Guarantees and limits of the correction]
\label{prop:correction}
Let an MPS $|\psi\rangle=JC$ be in canonical form with center $C$ at one of its endpoints after a
compression sweep, let $\hat O$ be Hermitian, and let $g$ be given by
Eq.~\eqref{eq:gradient} with $g \neq 0$. Then the following hold:

\emph{(i) Exact restoration.} If Eq.~\eqref{eq:quadratic} has a real solution $\mu$,
then $R_O(C+\mu g)=o_0$.

\emph{(ii) Rank preservation.} The tensors $C$ and $C+\mu g$ have the same shape.

\emph{(iii) Inertness in the absence of truncation.} If every timestep since the target
was recorded at $t_0$ discards no weight and its uncompressed update conserves
$\langle\hat O\rangle$, then $E=o_0$, the tolerance test of
Algorithm~\ref{alg:correction} returns before any displacement, and the
corrected step reproduces the uncorrected one exactly, provided optional norm
rescaling is absent or trivial.

\emph{(iv) Distance to the uncompressed state.} With $|\hat\psi\rangle$ the state before the
compression sweep, $|\psi\rangle=JC$ the state after it, and $X$ \emph{any}
tensor of the shape of $C$,
\begin{equation}\label{eq:pythagoras}
  \bigl\|J(C+X)-\hat\psi\bigr\|^2
  \;=\;
  \bigl\|\psi-\hat\psi\bigr\|^2 + \|X\|_\F^2.
\end{equation}
So every nonzero displacement of the center, the correction $X=\mu g$ included,
strictly increases the distance to the uncompressed state.
\end{proposition}

\begin{proof}
(i) With $\langle C,g\rangle_\F=0$ and $\Oeff$ Hermitian, the quantity
$\langle C+\mu g,\Oeff(C+\mu g)\rangle_\F-o_0\|C+\mu g\|_\F^2$ equals the left
side of Eq.~\eqref{eq:quadratic}, so any real root clears it, and $R_O$ is
defined because $\|C+\mu g\|_\F^2=\|C\|_\F^2+\mu^2\|g\|_\F^2>0$.

(ii) is immediate from Eq.~\eqref{eq:update}.

(iii) By induction on the steps. At $t_0$ the recorded value is $o_0$. If the
value entering a step is $o_0$, and that step discards no weight while its
uncompressed update conserves $\langle\hat O\rangle$, then $E=o_0$, the
tolerance test returns before the gradient is formed, the state is left
undisplaced, and the value entering the next step is again $o_0$.

(iv) By Lemma~\ref{lem:projection} the discarded component
$|\psi\rangle-|\hat\psi\rangle$ is orthogonal to every state of the form $JX$,
so in
$\|J(C+X)-\hat\psi\|^2
 =\|(\psi-\hat\psi)+JX\|^2$
the cross term $2\,\mathrm{Re}\langle JX,\psi-\hat\psi\rangle$ vanishes, and
$\|JX\|=\|X\|_\F$ holds by Eq.~\eqref{eq:isometry}.
\end{proof}

Clauses (i) to (iii) hold at every orthogonality center. The distance
identity~\eqref{eq:pythagoras} is guaranteed in the frame left by compression.
At another center the discarded component need not be orthogonal to the
available variations, so a correction can increase or decrease the distance
to $|\hat\psi\rangle$. Appendix~\ref{app:distance} gives the general cross-term
identity and its consequences.

Within the compression frame, a center displacement cannot recover the
discarded component. It can act only on the error already present within
that frame before compression; restoring an expectation value therefore
does not guarantee improved accuracy relative to the exact solution.

%% file: sections/admissibility.tex

\section{Sufficient conditions for projected-flow conservation}
\label{sec:admissibility}

To restore only compression-induced drift, the quantity must be an invariant
of the flow that the uncompressed updates
approximate. Let $\M_{\bm r}\subset\cH$ be the manifold of MPS with bond
dimensions $\bm r$~\cite{Holtz2012,Uschmajew2013}, and let $P_\psi$ be the
orthogonal projector onto its tangent space at $\psi$. Constraining $\dot\psi=-i\hat H\psi$ to $\M_{\bm r}$ by
the Dirac--Frenkel variational principle gives the projected flow
\begin{equation}\label{eq:diracfrenkel}
  \dot\psi \;=\; -i\,P_\psi\hat H\psi ,
\end{equation}
the dynamical low-rank approximation of the Schr\"odinger
equation~\cite{Lubich2014,Kieri2018}. Conservation by time-discrete and
rank-adaptive updates requires a separate analysis~\cite{Haegeman2016,Ceruti2022}.
The manifold is a
cone, since scaling one site tensor scales $|\psi\rangle$ and changes no bond
dimension. In particular the tangent space at $\psi$ contains $\psi$ itself, so
$P_\psi\psi=\psi$.

A quantity that is not an invariant of Eq.~\eqref{eq:diracfrenkel} drifts before
any weight is discarded, and pinning it after the compression fixes it to a
value that the low-rank dynamics does not maintain. Compatibility with this projected flow is therefore
characterized by the drift rate of $\langle\hat O\rangle$ along
Eq.~\eqref{eq:diracfrenkel}.

\begin{proposition}[Invariants of the projected flow]
\label{prop:invariants}
Let $\psi(t)$ solve Eq.~\eqref{eq:diracfrenkel} with $\|\psi(0)\|=1$, at bond
dimensions exactly $\bm r$, where $\M_{\bm r}$ is a smooth embedded
manifold~\cite{Holtz2012} and $P_\psi$ is defined. For every Hermitian $\hat O$
with $[\hat O,\hat H]=0$,
\begin{equation}\label{eq:driftrate}
  \frac{\dd}{\dd t}\langle\psi|\hat O|\psi\rangle
  \;=\; -2\,\mathrm{Im}\,\bigl\langle (\Id-P_\psi)\hat O\psi,\,
        (\Id-P_\psi)\hat H\psi\bigr\rangle .
\end{equation}
The rate vanishes wherever
\emph{(i)} $\hat O=\hat H$, or
\emph{(ii)} $\hat O\psi$ is tangent, $(\Id-P_\psi)\hat O\psi=0$.
Case~(ii) holds at every point of the manifold when
$\hat O=\sum_i\hat O_i$ is a sum of on-site operators, and at every eigenstate
of $\hat O$.
\end{proposition}

\begin{proof}
By $P_\psi\psi=\psi$ and the Hermiticity of $\hat H$ the quantity
$\langle\psi,P_\psi\hat H\psi\rangle=\langle\psi,\hat H\psi\rangle$ is real, so
the flow preserves the norm and the normalization of the expectation value may
be ignored. Differentiating and using $\hat O=\hat O^\dagger$,
\begin{equation*}
  \frac{\dd}{\dd t}\langle\psi|\hat O|\psi\rangle
  = 2\,\mathrm{Re}\,\langle\dot\psi,\hat O\psi\rangle
  = 2\,\mathrm{Im}\,\langle\hat O\psi,P_\psi\hat H\psi\rangle .
\end{equation*}
The product $\hat O\hat H$ is Hermitian because the two factors are Hermitian
and commute, so $\langle\hat O\psi,\hat H\psi\rangle=\langle\psi,\hat O\hat
H\psi\rangle$ is real and may be subtracted without changing the imaginary part,
leaving $-2\,\mathrm{Im}\langle\hat O\psi,(\Id-P_\psi)\hat H\psi\rangle$. Since
$\Id-P_\psi$ is an orthogonal projector it may be inserted into the first factor
as well, which gives Eq.~\eqref{eq:driftrate}.

In case~(i) the two factors coincide, the inner product is real, and the
imaginary part vanishes. In case~(ii) the first factor vanishes. An MPS is
multilinear in its site tensors, so for an on-site operator $\hat O_i$ the state
$\hat O_i\psi$ is the variation obtained from the single replacement
$A^{[i]}\mapsto\hat O_iA^{[i]}$ and is therefore tangent, and a sum of on-site
operators gives a sum of tangent vectors. For an eigenstate,
$\hat O\psi=\lambda\psi$ is tangent because the manifold is a cone.
\end{proof}

Energy, conserved on-site sums, and their real linear combinations form a
sufficient class of invariants of the fixed-rank projected flow;
Proposition~\ref{prop:invariants} does not establish an exhaustive classification.
Examples include total magnetization and particle number when their on-site
sums commute with $\hat H$, and conserved total-spin components. The generators
need not commute with one another. This agrees with the tangent-space
conservation condition in TDVP~\cite{Haegeman2011} and motivates the targets in
Sec.~\ref{sec:numerics}. Conservation by a particular uncompressed numerical
update requires a separate analysis. Other commuting observables, such as
$\hat S^2$, require an assessment using Eq.~\eqref{eq:driftrate}.

\subsection{Quantities supported on products across sites}
\label{sec:admissibility_s2}

The squared total spin $\hat S^2$ of an SU(2)-symmetric chain commutes with
$\hat H$ and is conserved by the exact flow. Its decomposition,
\[
  \hat S^2
  =\sum_i \hat{\mathbf S}_i^2
   +2\sum_{i<j}\hat{\mathbf S}_i\cdot\hat{\mathbf S}_j,
\]
contains cross-site terms, for which the single-site multilinearity argument
does not establish tangency. Cross-site structure alone does not imply drift:
the imaginary part in Eq.~\eqref{eq:driftrate} may vanish even when
$(\Id-P_\psi)\hat S^2\psi\neq0$. Whether $\langle\hat S^2\rangle$ is conserved
by the projected flow therefore depends on the Hamiltonian, the rank manifold,
and the state.

If the uncompressed update drifts, Algorithm~\ref{alg:correction} sees only
$|E-o_0|$ and cannot distinguish that drift from compression-induced loss.
It can therefore act on steps that discarded no weight, and the second
hypothesis of Proposition~\ref{prop:correction}(iii) then fails.

Because the gradients need not be mutually orthogonal (Sec.~\ref{sec:joint}),
a correction for $\hat S^2$ can also move other invariants.
Table~\ref{tab:excluded} quantifies this effect.

%% file: sections/joint.tex

\section{Several invariants at once}
\label{sec:joint}

Restoring two invariants by two scalar corrections generally fails. Each
displacement $\mu_ag_a$ can change other restored quantities at first order in
$\mu_a$, because the gradients $g_a$ need not be mutually orthogonal, so
corrections applied in sequence need not preserve earlier targets. Restoring the set
$\{\hat O_a\}_{a=1}^{p}$ with targets $o_a$ and post-compression values $E_a$
uses at each iteration the coupled displacement
\begin{equation}\label{eq:joint}
  C \;\longmapsto\; C + \sum_{a=1}^{p}\mu_a\,g_a,
  \qquad g_a=(O_{a,\mathrm{eff}}-E_a\Id)\,C ,
\end{equation}
with $O_{a,\mathrm{eff}}=J^\dagger\hat O_aJ$, $E_a=R_{O_a}(C)$, and the
gradients evaluated at the current $C$.
The multipliers are computed from the linearized $p$ constraints
$R_{O_a}\bigl(C+\sum_b\mu_bg_b\bigr)=o_a$ as described below.

Linearizing the constraints at $\bm\mu=0$ and using $\langle g_b,C\rangle_\F=0$
gives the Jacobian
\begin{equation}\label{eq:covariance}
  \mathcal{J}_{ab}
  \;=\; \frac{2\,\mathrm{Re}\,\langle g_a,\,g_b\rangle_\F}
             {\langle C,\,C\rangle_\F}
  \;=\; 2\,\Cov_C\!\bigl(O_{a,\mathrm{eff}},\,O_{b,\mathrm{eff}}\bigr),
\end{equation}
where $\Cov_C$ is the symmetrized covariance of the effective observables
in the normalized center $C/\|C\|_\F$.
The diagonal entries are twice the variances, and the linearized scalar
constraint is the $1\times1$ case. Forming the Jacobian requires no operator
applications beyond the $p$ gradients. As a real Gram matrix it is positive
semidefinite and singular exactly when the gradients are linearly dependent
over $\R$. This describes dependent first-order responses, not necessarily
redundant exact constraints or nonlinear infeasibility. The runs of
Sec.~\ref{sec:numerics} approach this degeneracy.

An undamped Newton step can be undefined or ill-conditioned there, so we use
a damped iteration with Levenberg--Marquardt scaling~\cite{Marquardt1963}:
\[
  (\mathcal J+\eta D)\bm\mu=-\bm\rho,
  \qquad \rho_a=R_{O_a}(C)-o_a.
\]
Here $\eta$ is the damping parameter and $D=\operatorname{diag}(\mathcal J)$, with zero entries replaced by its
largest entry. A trial is accepted when $\max_a|\rho_a|$ decreases, and the
gradients are recomputed at the new $C$. Numerical parameters are given in
Appendix~\ref{app:solver-settings}. Observables with negligible gradient are
dropped from the active iteration; this does not certify that their
constraints are satisfied or infeasible.

The displacement is confined to $C$, so rank preservation and, at the
compression center, the distance identity of Proposition~\ref{prop:correction}
apply. If every preceding step discards no weight and conserves all requested
targets, no correction is applied.
Joint restoration is complete only when all requested residuals meet
$\tau\max(1,\max_a|o_a|)$, including those of dropped observables; otherwise
it is unresolved. Each residual/gradient evaluation costs $p$ matrix-free
effective-operator applications, in addition to environment setup. Repeating
the joint solve at successive centers can improve convergence
(Sec.~\ref{sec:numerics}); the distance identity~\eqref{eq:pythagoras} remains
guaranteed only in the compression frame.

%% file: sections/numerics.tex

\section{Numerical demonstration}
\label{sec:numerics}

We test joint restoration with BUG and two-site TDVP, then examine scalar
correction failures and rank-preserving recovery. In both experiments, we compare
invariant restoration with accuracy against a full-state reference.

\subsection{Joint restoration}

We exercise the construction on the isotropic Heisenberg chain
\begin{equation}
    \hat H=\sum_{i=1}^{L-1}\hat{\mathbf S}_i\cdot\hat{\mathbf S}_{i+1} 
\end{equation}
for $L=16$. We write $\hat S^a=\sum_{i=1}^L\hat S_i^a$ ($a=x,y,z$)
and $\hat S^2=\sum_a(\hat S^a)^2$. The model
is SU(2)-symmetric, so it conserves the energy, every component of the total
spin, and $\hat S^2$. By Sec.~\ref{sec:admissibility} we first restore
$\{\hat H,\hat S^x,\hat S^y,\hat S^z\}$, four quantities, each a sum of
on-site operators or the generator itself, with $\hat S^2$ tested separately below. The joint
solve of Sec.~\ref{sec:joint} takes all four as one coupled system.

The initial state is the N\'eel product state with every spin rotated by $\pi/4$
about the $y$ axis. The N\'eel state itself is an eigenstate of $\hat S^z$, so its
$\hat S^z$ correction gradient vanishes by Eq.~\eqref{eq:gradient}. The rotation gives
every spin component nonzero variance, so truncation can perturb all of them.
The unpadded rotated state nevertheless obeys
$(\hat S^x+\hat S^z)|\psi_0\rangle=0$, giving a structural null direction in
the spin-gradient Gram matrix in that limit.
We take $T=6$, $\chi_{\max}=32$, $\Delta t=0.01$, and conservation
tolerance $10^{-13}$. Preparation and solver settings are given in
Appendix~\ref{app:experiment-settings}, together with the energy-only
controls, which required no scalar recovery. The implementation and
reproduction scripts are part of
YAQS~\footnote{\href{https://github.com/MaxFroehlich1410/yaqs/tree/91356ff463c03423e6cd2e608fee12a747c9f7fd}{Implementation and reproduction scripts, commit \texttt{91356ff}.}}.

The infidelity is $I(t)=1-|\langle\psi_{\mathrm{ref}}(t)|\psi(t)\rangle|^2$,
with both states normalized in this formula.

\begin{table}[tbp]
\caption{\label{tab:joint}%
\textbf{Joint restoration at $L=16$.}
Rotated N\'eel quench of the isotropic Heisenberg chain at $\chi_{\max}=32$,
evolved by the compressed BUG and by two-site TDVP, each without
correction and with the joint solve over
$\{\hat H,\hat S^x,\hat S^y,\hat S^z\}$.
Here $o_0^O$ denotes the initial expectation of $\hat O$, and $\delta_E=|\langle\hat H\rangle(T)-o_0^{H}|/|o_0^{H}|$ is the relative energy
drift, $\max_a\delta_{S^a}$ the largest absolute drift among the three
total-spin components, absolute because every $o_0^{S^a}$ vanishes here,
$\delta_{S^2}=|\langle\hat S^2\rangle(T)-o_0^{S^2}|$ the absolute drift of the total
spin, which is in neither solve, and $I$ the infidelity against the sparse
reference. All entries are at $T$.}
\footnotesize
\begin{ruledtabular}
\begin{tabular}{lcccc}
 & $\delta_E$ & $\max_a\delta_{S^a}$ & $\delta_{S^2}$ & $I$ \\
\hline
\multicolumn{5}{l}{BUG} \\
none  & $6.0\!\times\!10^{-5}$  & $6.2\!\times\!10^{-7}$
      & $8.0\!\times\!10^{-2}$  & $6.72\!\times\!10^{-2}$ \\
joint & $4.4\!\times\!10^{-15}$ & $3.5\!\times\!10^{-14}$
      & $1.0\!\times\!10^{-1}$  & $6.83\!\times\!10^{-2}$ \\
\hline
\multicolumn{5}{l}{2-TDVP} \\
none  & $1.1\!\times\!10^{-4}$  & $1.6\!\times\!10^{-13}$
      & $1.6\!\times\!10^{-1}$  & $7.03\!\times\!10^{-2}$ \\
joint & $4.7\!\times\!10^{-15}$ & $8.7\!\times\!10^{-17}$
      & $1.6\!\times\!10^{-1}$  & $7.04\!\times\!10^{-2}$ \\
\end{tabular}
\end{ruledtabular}
\end{table}

Both integrators restore every target to round-off at $T$
(Table~\ref{tab:joint}). The uncorrected two-site TDVP already holds the spin
components near round-off, whereas BUG compression perturbs them as well as
the energy. The joint correction changes the measured infidelity by at most
a few percent. It does not restore the untargeted $\hat S^2$ expectation, whose
drift can increase under a displacement made for the selected set.

\begin{table}[tbp]
\caption{\label{tab:excluded}%
\textbf{Adding $\hat S^2$.}
The BUG run of Table~\ref{tab:joint} with $\hat S^2$ added to the restored set, alone
with the energy and alongside the full admissible set. Each set is reported with
the correction applied at the one center the compression leaves and, in the row
below it, at every center in turn. Entries in bold are targets the solve met at
round-off at $T$. Here $\hat S^a$ abbreviates the three components
$\hat S^x,\hat S^y,\hat S^z$ and $\max_a\delta_{S^a}$ is the largest of their
drifts}
\footnotesize
\begin{ruledtabular}
\begin{tabular}{lcccc}
restored set & $\delta_E$ & $\max_a\delta_{S^a}$ & $\delta_{S^2}$ & $I$ \\
\hline
none & $6.0\!\times\!10^{-5}$ & $6.2\!\times\!10^{-7}$
     & $8.0\!\times\!10^{-2}$ & $6.72\!\times\!10^{-2}$ \\
$\{\hat H,\hat S^a\}$
     & $\bm{4.4\!\times\!10^{-15}}$ & $\bm{3.5\!\times\!10^{-14}}$
     & $1.0\!\times\!10^{-1}$  & $6.83\!\times\!10^{-2}$ \\
every center
     & $\bm{2.8\!\times\!10^{-15}}$ & $\bm{2.3\!\times\!10^{-14}}$
     & $8.6\!\times\!10^{-2}$  & $6.78\!\times\!10^{-2}$ \\
$\{\hat H,\hat S^2\}$
     & $\bm{6.2\!\times\!10^{-15}}$ & $3.2\!\times\!10^{-1}$
     & $\bm{1.8\!\times\!10^{-15}}$ & $1.53\!\times\!10^{-1}$ \\
every center
     & $\bm{2.5\!\times\!10^{-15}}$ & $3.5\!\times\!10^{-1}$
     & $\bm{3.2\!\times\!10^{-14}}$ & $1.85\!\times\!10^{-1}$ \\
$\{\hat H,\hat S^a,\hat S^2\}$
     & $4.9\!\times\!10^{-4}$ & $8.5\!\times\!10^{-4}$
     & $2.1\!\times\!10^{-2}$ & $1.26\!\times\!10^{-1}$ \\
every center
     & $\bm{4.9\!\times\!10^{-15}}$ & $\bm{7.0\!\times\!10^{-14}}$
     & $\bm{2.1\!\times\!10^{-14}}$ & $1.25\!\times\!10^{-1}$ \\
\end{tabular}
\end{ruledtabular}
\end{table}

\begin{figure}[tbp]
\includegraphics[width=\columnwidth]{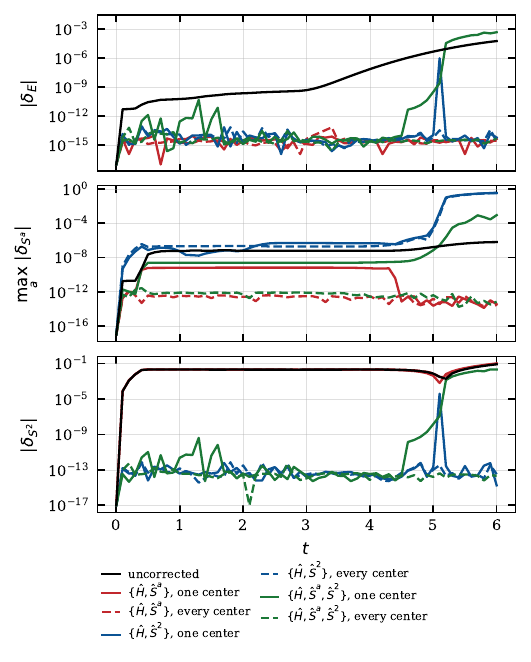}
\caption{\label{fig:trajectory}%
\textbf{The rows of Table~\ref{tab:excluded} resolved in time.}
}
\end{figure}

Table~\ref{tab:excluded} shows two different effects of adding $\hat S^2$.
Restoring $\{\hat H,\hat S^2\}$ succeeds, but increases the spin-component
errors and more than doubles the infidelity. With all five quantities at one
center, nearly dependent gradients make the Gram matrix ill-conditioned and
the iteration fails to meet the targets. Repeating the solve at successive
centers changes the available directions and restores all five targets at
$T$, while the infidelity remains about twice the uncorrected value.

The trajectories in Fig.~\ref{fig:trajectory} also distinguish the schedules
when their final target errors agree. For the four-target solve at one
center, the spin error remains on a plateau for roughly the first two thirds
of the run while the gradients are nearly dependent. As they separate, the
residual falls to round-off. The every-center solve avoids this stagnation
and stays about three orders below the plateau.

Restoring global expectations does not fix their spatial distribution.
For the every-center runs, adding $\hat S^2$ to the four-target set raises the
maximum local-spin error from $8.7\times10^{-4}$ to $5.8\times10^{-2}$ and
the staggered-magnetization error from $2.5\times10^{-4}$ to $1.1\times10^{-2}$,
while bond-energy and correlator errors change little
(Table~\ref{tab:observables}, Appendix~\ref{app:local-errors}). The local
errors can cancel in the conserved total, so exact restoration alone does
not certify dynamical accuracy.

\input{subsection_fallback}

%% file: subsection_fallback.tex
\subsection{Scalar failure recovery}
\label{sec:scalar-fallback}

The scalar correction can fail at the center left by compression even when
the compressed trajectory is accurate. We test the recovery of Sec.~\ref{sec:correction_recovery}
on open dimerized Heisenberg chains,
\begin{equation}
  \hat H_r=\sum_{j=1}^{L-1}J_j\hat{\mathbf S}_j\cdot\hat{\mathbf S}_{j+1},
  \qquad
  J_j=\begin{cases}1,&j\text{ odd},\\r,&j\text{ even},\end{cases}
\end{equation}
quenched from a product of singlets on the strong bonds. We take
$r=0.05,0.2,0.5$, $L=12,16$, $\chi_{\max}=16,32,64$,
$\Delta t=0.02,0.01$, and $T=8$. A domain-wall quench of an isotropic
inverse-square ring provides an additional control. Each Hamiltonian is
scaled to spectral bandwidth per site one, without shifting its energy
origin; further preparation details are in Appendix~\ref{app:experiment-settings}.

All runs use BUG with conservation tolerance $\tau=10^{-13}$. We compare
no correction, scalar correction alone, and scalar correction with recovery.
BUG calls the correction twice per timestep; counts include calls whose
energy already meets tolerance. The full-state reference is evaluated at
the actual correction time, including half steps.
In addition to infidelity $I$, we measure
\begin{equation}
  \epsilon_{\mathrm{obs}}(t)
  =\max_{\hat A\in\mathcal A}
  \bigl|\langle\hat A\rangle_{\psi(t)}-\langle\hat A\rangle_{\psi_{\mathrm{ref}}(t)}\bigr|,
\end{equation}
where $\mathcal A$ contains all local spin components
$\hat S_i^a$ and nearest-neighbor equal-axis correlations
$\hat S_i^a\hat S_{i+1}^a$, $a=x,y,z$.
A call is accuracy-qualified when $I\leq10^{-3}$ and
$\epsilon_{\mathrm{obs}}\leq10^{-3}$ before correction.

\input{table_fallback_counts}
\input{figure_fallback}

Table~\ref{tab:fallback-counts} reports the weak-dimer runs, all of whose
correction calls are accuracy-qualified. The $271$ recoveries occur only at
$\chi_{\max}=16$ for $r=0.05,0.2$; halving the timestep does not eliminate
them. Every event is a negative discriminant, resolved at the first other
center visited. The full-frame stage is never reached, so these data test
center scanning rather than the full-frame minimizer. The $r=0.5$ and
inverse-square controls require no recovery; their detailed counts are
given in Appendix~\ref{app:experiment-settings}. Absence of recovery is not
an accuracy test.

For the representative $r=0.2$, $L=16$, cap-$16$ run at $\Delta t=0.01$
(Fig.~\ref{fig:scalar-fallback}), recovery reduces the maximum absolute
energy error from $6.49\times10^{-9}$ for the scalar correction alone to
$1.19\times10^{-12}$, compared with $2.03\times10^{-6}$ without correction.
All three trajectories remain within the accuracy thresholds, but the
maximum observable error increases from $4.61\times10^{-4}$ to
$5.36\times10^{-4}$ with recovery. The gain is reliable energy restoration,
not uniform error reduction; corrections at another center also lack the
guaranteed compression-frame orthogonality of
Proposition~\ref{prop:correction}(iv). The frequencies refer to this fixed-seed
grid and conservation tolerance, not to general occurrence probabilities.

%% file: table_fallback_counts.tex
\begin{table}[tbp]
\caption{\textbf{Scalar fallback in weak-dimer quenches.}
BUG evolution from a singlet product to $T=8$, with spectral bandwidth per site equal to one.
Entries give fallback attempts per scalar correction call, $F/N$, including calls that require no correction.
Every call in the displayed runs is accuracy-qualified ($I,\epsilon_{\mathrm{obs}}\leq10^{-3}$ before correction), so $F/N=F_{\mathrm{acc}}/N_{\mathrm{acc}}$.
All attempts succeeded by center scanning; the full-frame stage was never reached.
The $r=0.5$ and inverse-square zero-event controls are described in Appendix~\ref{app:experiment-settings}.}
\label{tab:fallback-counts}
\footnotesize
\begin{ruledtabular}
\begin{tabular}{cccrrr}
 & & & \multicolumn{3}{c}{$F/N=F_{\mathrm{acc}}/N_{\mathrm{acc}}$} \\
$r$ & $L$ & $\Delta t$ & $\chi_{\max}=16$ & $\chi_{\max}=32$ & $\chi_{\max}=64$ \\
\hline
0.05 & 12 & 0.02 & $12/800$ & $0/800$ & $0/800$ \\
 & 12 & 0.01 & $10/1600$ & $0/1600$ & $0/1600$ \\
 & 16 & 0.02 & $18/800$ & $0/800$ & $0/800$ \\
 & 16 & 0.01 & $26/1600$ & $0/1600$ & $0/1600$ \\
\hline
0.2 & 12 & 0.02 & $21/800$ & $0/800$ & $0/800$ \\
 & 12 & 0.01 & $53/1600$ & $0/1600$ & $0/1600$ \\
 & 16 & 0.02 & $38/800$ & $0/800$ & $0/800$ \\
 & 16 & 0.01 & $93/1600$ & $0/1600$ & $0/1600$ \\
\end{tabular}
\end{ruledtabular}
\end{table}

%% file: figure_fallback.tex
\begin{figure}[tbp]
\includegraphics[width=\columnwidth]{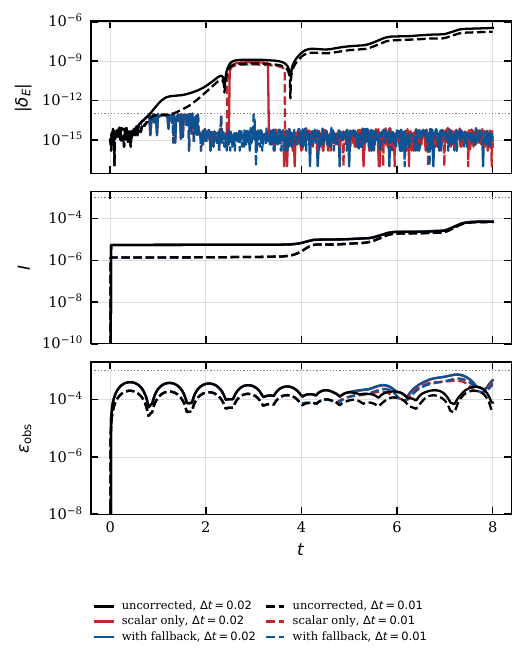}
\caption{\textbf{Scalar failure recovery during accurate evolution.}
Dimer quench at $r=0.2$, $L=16$, and $\chi_{\max}=16$, with
$\Delta t=0.02$ (solid) and $0.01$ (dashed).
Black is uncorrected BUG, red the scalar correction alone, and blue the
correction with fallback. The panels show the relative energy drift
$|\delta_E|$ of Table~\ref{tab:joint}, infidelity $I$, and maximum spin-observable error
$\epsilon_{\mathrm{obs}}$ against the full-state reference.
Dotted gray lines mark the conservation tolerance $10^{-13}$ and the two
accuracy thresholds $10^{-3}$. Curves are sampled at completed timesteps;
the counts in Table~\ref{tab:fallback-counts} also include half-step calls.
Recovery counts are given in Table~\ref{tab:fallback-counts}.
Values below $10^{-17}$ are floored for logarithmic display.}
\label{fig:scalar-fallback}
\end{figure}

%% file: sections/conclusion.tex

\section{Conclusion}
\label{sec:conclusion}

We have developed a post-compression correction that restores selected
expectation values within a canonical MPS without increasing its bond
dimensions. For one observable, the exact constraint reduces to a scalar
quadratic, with restoration possible along the chosen direction whenever a
real root exists. Rank-preserving recovery searches other centers and, if
needed, the complete one-site frame; a coupled solve treats several
observables. Numerical experiments demonstrate joint restoration for BUG and
two-site TDVP, and scalar recovery by center scanning along accurate BUG
trajectories.

This construction connects geometric numerical integration with the canonical
geometry of MPS. It makes conservation an explicit design criterion within
the compressed representation and separates three questions: what the
projected flow conserves, what the numerical update conserves, and what can
be restored after compression. These distinctions provide a framework for
choosing targets and assessing rank-adaptive algorithms through conservation,
approximation error, and computational cost.

The benefits remain conditional on the target and the available correction
space. Energy and conserved on-site sums belong to a sufficient projected-flow
invariant class; other commuting quantities, such as $\hat S^2$, require
individual assessment. Within the compression frame, restoration cannot
recover discarded information, and the experiments show that improved
conservation need not improve state or observable accuracy. Future work
should assess these trade-offs across a broader range of models and
integrators, including scalar recovery with two-site TDVP, and test the
robustness and cost of full-frame recovery, which was not required in the
present experiments.

%% file: appendices.tex

\section{Additional fixed-frame distance identities}
\label{app:distance}

Clauses (i) to (iii) of Proposition~\ref{prop:correction} hold at every
orthogonality center. Clause~(iv) is the case
$C=J^\dagger\hat\psi$ of
\begin{equation}\label{eq:displacement}
\begin{split}
  \bigl\|J(C+X)-\hat\psi\bigr\|^2
  &\;=\;
  \bigl\|\psi-\hat\psi\bigr\|^2 + \|X\|_\F^2\\
  &\quad + 2\,\mathrm{Re}\,\langle X,C-J^\dagger\hat\psi\rangle_\F ,
\end{split}
\end{equation}
which holds for any isometry $J$ and any tensor $X$ of the shape of $C$.
Lemma~\ref{lem:projection} gives $C=J^\dagger\hat\psi$ for the embedding the compression
sweep produced, and Algorithm~\ref{alg:correction} applies the correction in
that gauge. At a center the sweep did not leave, $C-J^\dagger\hat\psi$ is in general nonzero
and the cross term carries either sign, so a displacement can reduce the
distance to $|\hat\psi\rangle$ as well as increase it. The reduction is largest
at $X=J^\dagger\hat\psi-C$ and equals $\|C-J^\dagger\hat\psi\|_\F^2$.

At the center the sweep leaves, the orthogonality of
Lemma~\ref{lem:projection} also bounds the effect of the correction on the
accuracy. Let $|\psi_{\mathrm{ex}}(t)\rangle$ denote the exact solution. Since
$|\psi\rangle-|\hat\psi\rangle$ is orthogonal to every state of the form $JX$,
\begin{equation}\label{eq:blind}
  \langle JX|\psi-\psi_{\mathrm{ex}}(t)\rangle \;=\; \langle JX|\hat\psi-\psi_{\mathrm{ex}}(t)\rangle
\end{equation}
for every tensor $X$ of the shape of $C$. Expanding
\begin{equation*}
\begin{split}
  \bigl\|J(C+X)-\psi_{\mathrm{ex}}(t)\bigr\|^2
  &= \bigl\|\psi-\psi_{\mathrm{ex}}(t)\bigr\|^2 + \|X\|_\F^2\\
  &\quad + 2\,\mathrm{Re}\,\langle JX,\,\psi-\psi_{\mathrm{ex}}(t)\rangle
\end{split}
\end{equation*}
leaves the cross term as the only one a displacement can make negative, and by
Eq.~\eqref{eq:blind} that term is the same as it was before the compression. The
correction can therefore work only against the error the state already carried,
and the weight the compression has just discarded is invisible to it.

\section{Numerical implementation and additional diagnostics}
\label{app:implementation}

\subsection{Joint-solver parameters}
\label{app:solver-settings}

The joint solve starts from $\eta=10^{-3}$. Damping is halved after
acceptance, with floor $10^{-12}$, and quadrupled after rejection, with at most
12 trials per iteration and 30 iterations. An observable is dropped from
the active solve when $\|g_a\|_\F^2<10^{-28}$. Dropped constraints remain
included in the final residual test of Sec.~\ref{sec:joint}.

\subsection{Preparation, tolerances, and controls}
\label{app:experiment-settings}

For the joint-restoration experiment of Sec.~\ref{sec:numerics},
we take $T=6$, $\chi_{\max}=32$, and $\Delta t=0.01$, with relative
discarded-weight threshold $10^{-12}$, Krylov tolerance $10^{-12}$ and
conservation tolerance $10^{-13}$. Initial tensors use $10^{-10}$ padding to
maximum bond dimension four (seed 20260812).

Corresponding energy-only controls, with the same Heisenberg setup, required
no scalar fallback: there were $0/1200$ attempts per correction call for BUG
and $0/600$ for two-site TDVP. Restricting the denominators to the
accuracy-qualified calls defined in Sec.~\ref{sec:scalar-fallback} gives $0/914$ and $0/458$,
respectively. Enabling fallback left each scalar-only trajectory unchanged;
these controls are separate from the joint restoration in
Tables~\ref{tab:joint} and~\ref{tab:excluded}.

For the scalar-recovery experiment, the dimer quenches use
$r=0.05,0.2,0.5$, $L=12,16$, $\chi_{\max}=16,32,64$,
$\Delta t=0.02,0.01$, and $T=8$. The inverse-square ring has couplings
$d_{ij}^{-2}$ with $d_{ij}=(L/\pi)\sin(\pi|i-j|/L)$ and a domain-wall initial
state. Hamiltonians are scaled so that $(E_{\max}-E_{\min})/L=1$ without
shifting the energy origin. This differs from the unscaled joint-restoration
experiment. Prepared initial tensors and Hamiltonian scales are shared across
comparison arms and refinements. Singlet preparations use seeded rank-two
bond padding of order $10^{-10}$.

Scalar-recovery runs use a hard bond cap, zero SVD threshold, Krylov tolerance
$10^{-14}$, and conservation tolerance $10^{-13}$. There are $800$ correction
calls at $\Delta t=0.02$ and $1600$ at $\Delta t=0.01$, including calls
already within tolerance. The main-text accuracy criteria are evaluated
before correction against the full-state reference at the actual correction
time, including half steps.

The $r=0.5$ dimer and inverse-square controls required no fallback at
either size or timestep and any of the three caps: $0/14400$ calls for each
Hamiltonian, or $0/9509$ and $0/2752$ accuracy-qualified calls, respectively.
Only a subset of these control calls is accuracy-qualified. For $L=12$,
cap $64$ is full rank; it removes truncation but leaves time-integration error.

\subsection{Errors in unconstrained observables}
\label{app:local-errors}

Table~\ref{tab:observables} gives the full comparison of local-spin,
bond-energy, correlator, and staggered-magnetization errors discussed in
Sec.~\ref{sec:numerics}. The data distinguish restoration of a global sum
from accuracy of the local expectations that form it.

\par\medskip
\noindent\begin{minipage}{\columnwidth}
\makeatletter\def\@captype{table}\makeatother
\caption{\label{tab:observables}%
\textbf{Errors in unconstrained observables.}
The uncorrected run and the two every-center arms of Table~\ref{tab:excluded}.
Every entry is the largest absolute deviation from the exact solution at $T$,
over the $16$ sites and three axes of the local spin, over the $15$ bonds, and
over the seven separations $j$ of the correlator, which is taken from the chain
center $c=L/2+1$. The staggered magnetization
$M_s=L^{-1}\sum_i(-1)^i\langle\hat S^z_i\rangle$ is a single number.}
\footnotesize
\begin{ruledtabular}
\setlength{\tabcolsep}{2pt}
\begin{tabular}{lcccc}
restored set & local spin & bond energy & correlator & staggered \\
 & $\langle\hat S^a_i\rangle$
 & $\langle\hat{\mathbf S}_i\!\cdot\!\hat{\mathbf S}_{i+1}\rangle$
 & $\langle\hat S^z_c\hat S^z_{c+j}\rangle$
 & $M_s$ \\
\hline
none
 & $8.7\!\times\!10^{-4}$ & $1.22\!\times\!10^{-2}$ & $2.1\!\times\!10^{-3}$
 & $6.0\!\times\!10^{-5}$ \\
$\{\hat H,\hat S^a\}$
 & $8.7\!\times\!10^{-4}$ & $1.23\!\times\!10^{-2}$ & $2.0\!\times\!10^{-3}$
 & $2.5\!\times\!10^{-4}$ \\
$\{\hat H,\hat S^a,\hat S^2\}$
 & $5.8\!\times\!10^{-2}$ & $1.23\!\times\!10^{-2}$ & $2.4\!\times\!10^{-3}$
 & $1.1\!\times\!10^{-2}$ \\
\end{tabular}
\end{ruledtabular}
\end{minipage}\par\medskip